\documentclass[%
 aip,
 sd,%
 amsmath,amssymb,
 reprint,%
]{revtex4-1}
\usepackage{color}
\usepackage{graphicx}
\usepackage{dcolumn}
\usepackage{bm}
\pdfoutput=1
\begin{document}
\title{Charged particle dynamics in the presence of non-Gaussian L\'{e}vy electrostatic fluctuations}
\author{Sara Moradi$^{1}$, Diego del-Castillo-Negrete$^{2}$, and Johan Anderson$^3$}
  \address{$^{1}$ Fluid and Plasma Dynamics, UniversiteŽ Libre de Bruxelles, 1050 Brussels, Belgium \\
  $^{2}$ Oak Ridge National Laboratory, Oak Ridge, Tennessee 37831-8071, USA\\
  $^{3}$ Department of Earth and Space Sciences, Chalmers University of Technology, SE-412 96 G\"{o}teborg, Sweden}

\begin{abstract}
Full orbit dynamics of charged particles in a $3$-dimensional helical magnetic field in the presence of $\alpha$-stable L\'{e}vy electrostatic fluctuations and linear friction modeling collisional Coulomb drag is studied via Monte Carlo numerical simulations. The  L\'{e}vy fluctuations are introduced to model the effect of non-local transport due to fractional diffusion in velocity space resulting from intermittent electrostatic turbulence. The probability distribution functions of energy, particle displacements, and Larmor radii are computed and showed to exhibit a transition from exponential decay, in the case of Gaussian fluctuations, to  power law decay in the case of L\'{e}vy fluctuations. The absolute value of the power law decay exponents are linearly proportional to the L\'{e}vy index $\alpha$. The observed anomalous non-Gaussian statistics of the particles' Larmor radii (resulting from outlier transport events) indicate that, when electrostatic turbulent fluctuations exhibit non-Gaussian L\'{e}vy statistics, gyro-averaging and guiding centre approximations might face limitations and full particle orbit effects should be taken into account.   
\end{abstract}
\maketitle
There is a considerable amount of experimental evidence \cite{Lopez,Gentle,Callen,Mantica,van-Milligen,BalescuBook,del-castillo2008} and numerical gyrokinetic \cite{Dif-Pradalier,Sanchez} and fluid \cite{del-Castillo-Negrete2005} simulations that indicate that plasma turbulent transport in tokamaks is, under some conditions, non-diffusive. There are several reasons for the possible breakdown of the standard diffusion paradigm which is based on restrictive assumptions including locality, Gaussianity, lack of long range correlations, and linearity. Different physical mechanisms can generate situations where e.g., locality and Gaussianity may be incorrect assumptions for understanding transport. For example, interactions with external fluctuations may introduce long-range correlations and/or anomalously large particle displacements. The source of the external fluctuations could  be that not all relevant physics is taken into account such as coherent modes or other non-linear mechanisms. The emergence of such strange kinetics has been studied previously, e.g., \cite{schlesinger1993, metzler2000, metzler2004,chechkin2002,del-Castillo-Negrete2004,del-Castillo-Negrete2010,anderson} using different modeling strategies where it may be generated by accelerated or sticky motions along the trajectory of the random walk. 

In addition, turbulence intermittency is characterized by patchy spatial structures that are bursty in time and coupling to these modes introduces long range correlations and/or L\'{e}vy distributed noise characteristics. The probability density functions (PDF) of intermittent events often show unimodal structure with ``elevated" tails that deviate from Gaussian predictions \cite{Carreras96,Carreras99,Jha2003}. Experimental evidence of L\'{e}vy statistics in the electrostatic fluctuation at the plasma edge was presented in Ref.~\cite{Jha2003}, with a L\'{e}vy index in the range $\alpha = 1.1 - 1.3$  at short times and in the range $\alpha = 1.8 - 2$  at long times. Furthermore, in Ref. \onlinecite{Mizuuchi2005} it was observed that moving from the inner to the outer region of edge plasma, the L\'{e}vy index decreases, suggesting that the PDFs of the turbulence near the boundary region of Heliotron J are nearly Gaussian, whereas at the outer regions of plasma they become strongly non-Gaussian. The statistics of the measured fluctuations at the edge of Stellarators such as ÒUragan 3MÓ and HELIOTRON J have been observed to change from L\'{e}vy to Gaussian at the L to H-mode transition \cite{Gonchar2003,Burnecki 2012,Burnecki 2015}. These type of observations are not limited to fusion plasmas, L\'{e}vy-type turbulent random processes and related anomalous diffusion phenomena have been observed in a wide variety of complex systems such as semiconductors, glassy materials, nano-pores, biological cells, and epidemic spreading \cite{Bramwell}. The kinetic descriptions which arise as a consequence of averaging over the well-known Gaussian statistics seem to fall short in describing the apparent randomness of these dynamical chaotic systems. Thus, the problem of finding a proper kinetic description for such complex systems is a challenge. 

L\'{e}vy statistics \cite{levy} describing fractal processes (L\'{e}vy index $\alpha$ where $0 < \alpha < 2$ ) lie at the heart of complex processes such as anomalous diffusion \cite{metzler2000}. L\'{e}vy statistics can be generated by random processes that are scale-invariant with anomalous scaling exponents. This means that a trajectory lacks a unique characteristic scale that dominates the process. Geometrically this implies the fractal property that a trajectory, viewed at different resolutions, will exhibit self-similar properties. Indeed, self-similar analysis of fluctuation measurements by Langmuir probes in different fusion devices such as spherical tokamak, reversed field pinch, stellarator, and several tokamaks, have provided evidence to support the idea that density and potential fluctuations are distributed according to L\'{e}vy statistics \cite{Carreras99}. Furthermore, the experimental evidence of the wave-number spectrum characterised by power laws over a wide range of wave-numbers can be directly linked to the values of L\'{e}vy index $\alpha$ of the PDFs of the underlying turbulent processes. 

In a previous study \cite{anderson} the aim was to shed light on the non-extensive properties of the velocity space statistics and characterization of the fractal processes limited to the Fractional Fokker-Planck Equation in terms of Tsallis statistics. The goal of this paper is to study the statistics of charged particle motion in the presence of $\alpha$-stable L\'{e}vy fluctuations in a external magnetic field and linear friction using  Monte Carlo numerical simulations. The L\'{e}vy noise is introduced to model the effect of non-Gaussian, intermittent electrostatic fluctuations. The statistical properties of the velocity moments and energy for various values of the L\'{e}vy index $\alpha$ are investigated as well as the role of L\'{e}vy fluctuations on the statistics of the particles' Larmor radii in order to examine potential limitations of gyro-averaging. Fractional kinetics of charged particle transport in  a constant parallel magnetic field and a random electric field was studied in Ref. \cite{chechkin2002}. Going beyond this work, we perform $3$-dimensional simulations  in a helical magnetic field  and study the statistics of the spatial displacements and Larmor radius which were not discussed in Ref. \cite{chechkin2002} whose numerical results were limited to $2$-dimensions using a different type of isotropic L\'{e}vy processes. However, memory effects are neglected since the L\'{e}vy noise is taken as white or delta correlated in time.

We consider the motion of charged particles in a $3$-dimensional magnetic field in a cylindrical domain in the presence of  linear friction modeling collisional Coulomb drag and a stochastic electric field according to the Langevin equations
\begin{eqnarray}
&&\frac{d {\bf r}}{dt}= {\bf v},\label{eqr}\\
&&\frac{d {\bf v}}{dt}= \frac{q_s}{m_s} {\bf v} \times {\bf B} -\nu {\bf v} +\frac{q_s}{m_s} \mathcal{E},\label{eqv}
\end{eqnarray}
where $q_s$ and $m_s$ are the charge and mass of the particle species $s$, $\nu$ is the friction parameter and $\mathcal{E}$ is a 3-dimensional, homogeneous, isotropic turbulent electric field modeled as an stationary, uncorrelated stochastic process without memory following an $\alpha$-stable distribution, $f(\alpha,\beta,\sigma,\eta)$, with characteristic exponent $0<\alpha\le 2$, skewness $\beta=0$, variance $\sigma=1/\sqrt{2}$, and mean $\eta=0$. Here, we use the definition of $f(\alpha,\beta,\sigma,\eta)$ as described in Refs. \cite{Borak,Chambers,Weron}.

A periodic straight cylindrical domain with period $L=2 \pi R_0$ is considered, with $R_{0}$ being the major radius, and we use cylindrical coordinates $(r, \theta,z)$. The magnetic field is a helical field of the form, 
\begin{equation}
\label{B_model}
{\bf B}(r)=B_\theta(r)\,  \hat{\bf e}_\theta + B_z \hat{\bf e}_z.
\end{equation}
A constant magnetic field in $z$-direction, $B_z=B_0$, is assumed. The shear of the helical magnetic field, i.e. the dependence of the azimuthal rotation of the field as function of the radius, is determined by the $q$-profile, $q(r)=r B_z/(R_0 B_\theta)$, where 
\begin{equation}
\label{B_theta}
B_\theta(r)=\frac{B (r /\lambda)}{1+(r/\lambda)^2} \, ,
\end{equation}
for which the $q$ profile is
\begin{equation}
\label{q_monotonic}
q(r)=q_0\left( 1 + \frac{r^2}{\lambda^2} \right ) \, .
\end{equation}
In terms of the flux variable,
\begin{equation}
\label{psi_def}
\psi=\frac{r^2}{2 R_0^2} \, ,
\end{equation} 
$q$ is a linear function of $\psi$. 

The numerical integration of Eqs. (\ref{eqr}) and (\ref{eqv}) is performed using a Runge-Kutta 4th order scheme (RK4) over the interval $[0,T]$. The time step for the RK4 integration is defined by partitioning the interval $[0,T]$ into $N$ subintervals of width $\delta=T/N>0$,
\begin{equation}
0 = \tau_{0} < \tau_{1} < \dots < \tau_{i} < \tau_{N} = T \, ,
\end{equation}
with the initial conditions $\mathbf{r}_0,$ and $\mathbf{v}_0$. We compute $\mathbf{r}_i$ and $\mathbf{v}_{i}$ for the subintervals with the time step of $dt = \delta/n$, and at every $\delta$, we include the cumulative integral of the stochastic process using  
\begin{eqnarray}
&&d {\bf r}_i= {\bf v}_i dt\\
&&d {\bf v}_i= \left[\frac{q_s}{m_s} {\bf v}_i \times {\bf B} -\nu {\bf v}_i \right] dt  + {\bf W}
\end{eqnarray}
where
\begin{equation}
{\bf W} =  \frac{q_s}{m_s} \chi \sum_{\delta} (dt)^{(1/\alpha)} {\bf \mathcal{E}}.
\end{equation}
Here, using spherical coordinates, random samples in the $\mathcal{E}_{\rho}$ radial direction are generated with the $\alpha$-stable random generator developed in Ref. \cite{Borak,Chambers,Weron}, and two uniformly distributed angles $\theta$ and $\phi$ between $[0, 2\pi]$ are used. In Cartesian coordinates the components of the electric field are $\mathcal{E}_x=\mathcal{E}_{\rho} \sin \theta \cos \phi$, $\mathcal{E}_y=\mathcal{E}_{\rho} \sin \theta \sin \phi$, and  $\mathcal{E}_z=\mathcal{E}_{\rho} \cos \theta$. $N_p=10^4$ particles are considered, and the simulation time is $T=500/\tau_c$ where $\tau_c=2\pi/\Omega_c$ and $\Omega_c=|q_s| B_0/m_s$ is the gyration frequency. We explore the dependence of the particle motion on the index $\alpha$ of the L\'{e}vy fluctuations and the parameter $\epsilon=\chi/\nu$ where $\chi$ is the amplitude of the fluctuations and $\nu$ is the damping coefficient. The convergence in probability of L\'{e}vy driven stochastic differential equations \ref{eqr} and \ref{eqv} have been discussed in \cite{Pavlyukevich2014} where a criteria is established.

\begin{figure}
\includegraphics[width=6cm, height=4.cm]{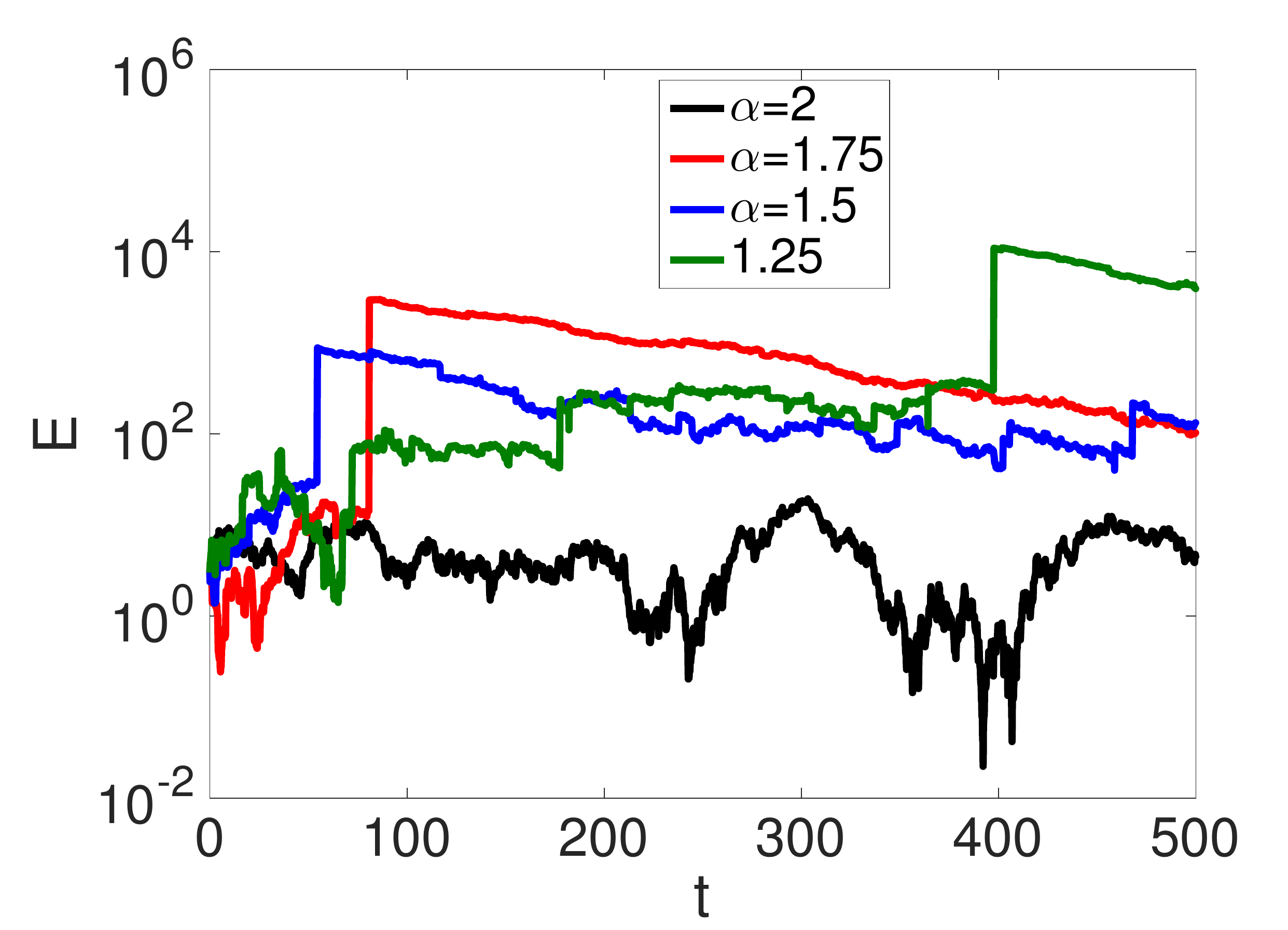} 
\caption{\label{fig4} Samples of normalised particle energy, $E = \frac{1}{2}(v_x^2 + v_y^2 +v_z^2)/v^2(0)$, vs time for different values of $\alpha=2$ (black), $1.75$ (red), $1.5$ (blue), and $1.25$ (green). Here, $\epsilon=100$.}
\end{figure}

\begin{figure}
\includegraphics[width=6cm, height=4.cm]{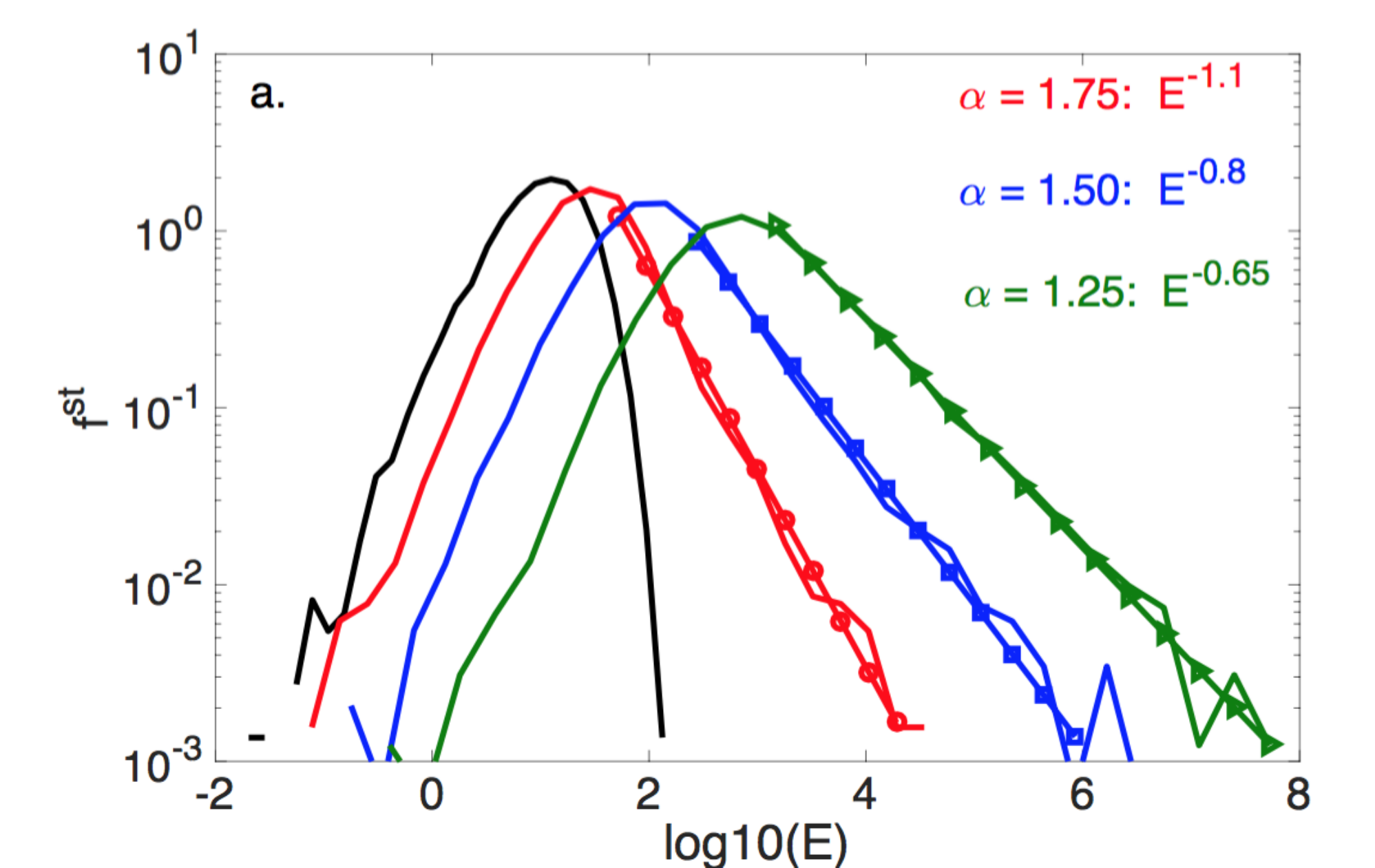} \\
\includegraphics[width=6cm, height=4.cm]{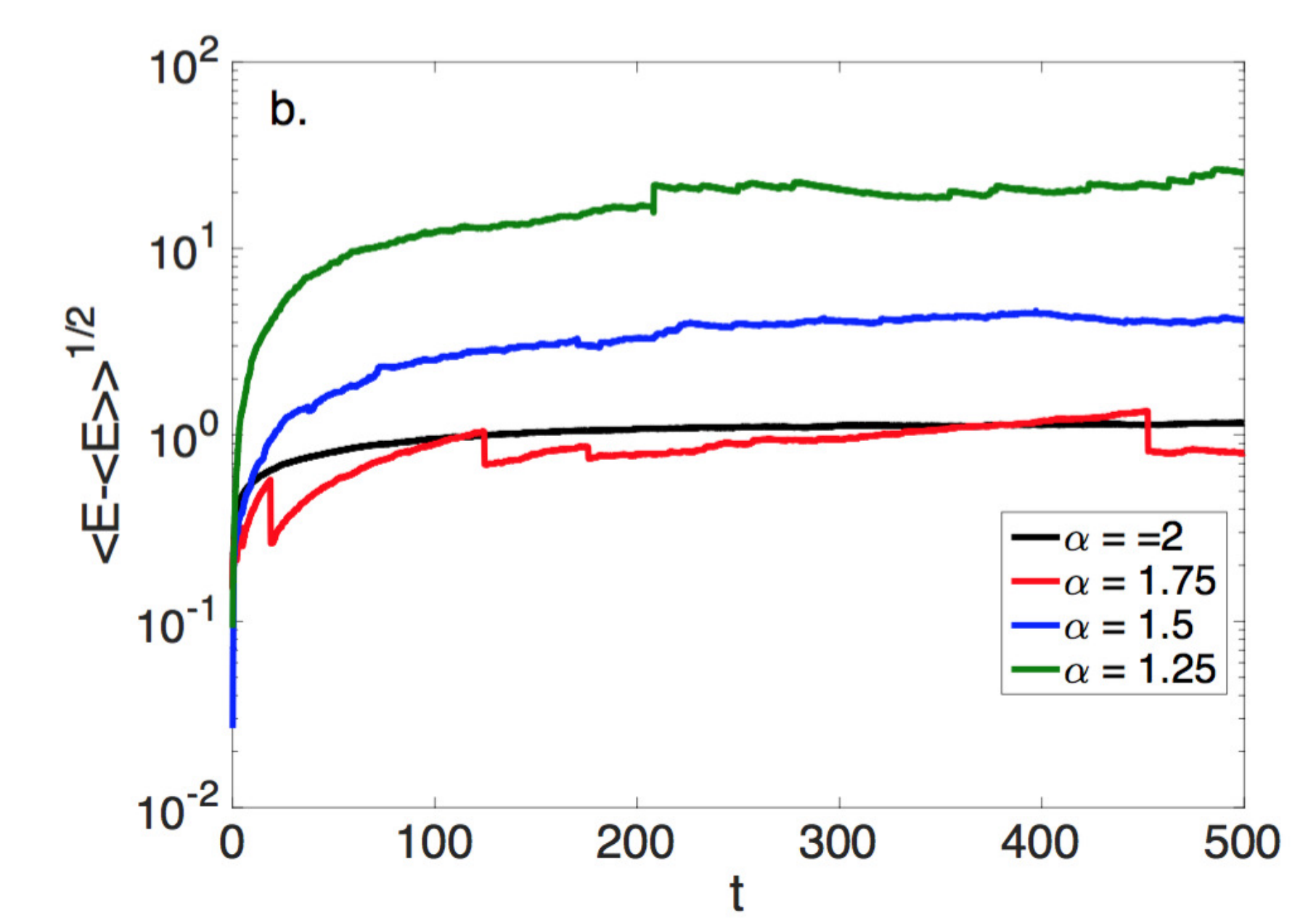} 
\caption{\label{fig5} (a) The steady state PDFs ($f^{st}$) of $Log_{10}(E)$, and linear fits for the energy decay are shown by lines with symbols, (b) the $q$-moment are shown with $q=1/2$ for different values of $\alpha=2$ (black), $1.75$ (red), $1.5$ (blue), and $1.25$ (green). Here, $\epsilon=100$.}
\end{figure}

Figure~\ref{fig4} shows samples of the particles' energy as function of time for several values of $\alpha$. It is observed that as $\alpha$ decreases, the random walk in energy is strongly influenced by outlier events which result in intermittent behavior with appearance of L\'{e}vy flights between periods of small perturbations. The rate and the amplitude of the intermittent jumps in energy increase significantly as $\alpha$ is decreased. This behavior is clearly observed in Figs. \ref{fig5}(a) and (b) where the PDF of $Log_{10}$ of the particle energy, $E$, and the $q=1/2$-moment \cite{anderson} of the energy as functions of time are shown. As seen in Fig. \ref{fig5}(a) the decay of the PDFs changes from exponential in the case of a Gaussian process to power law in the case of  a L\'{e}vy process. The power law exponent decreases as $\alpha$ is decreased indicating the increase in the probability of the occurrence of L\'{e}vy flights. A breakup in the symmetry of the PDFs is also observed with a shift towards higher values of the energy, as the L\'{e}vy index $\alpha$ is reduced. Note that the numerical results indicate that the PDFs relax towards stationary states. The $q=1/2$-moments of the energy converge in the considered simulation time span, and there exist about two orders of magnitude increase in the converged values as $\alpha$ varies from a Gaussian process ($\alpha=2$) towards a strongly L\'{e}vy distributed process ($\alpha=1.25$), as can be seen in Fig. \ref{fig5}(b).

\begin{figure}
\includegraphics[width=6cm, height=4.cm]{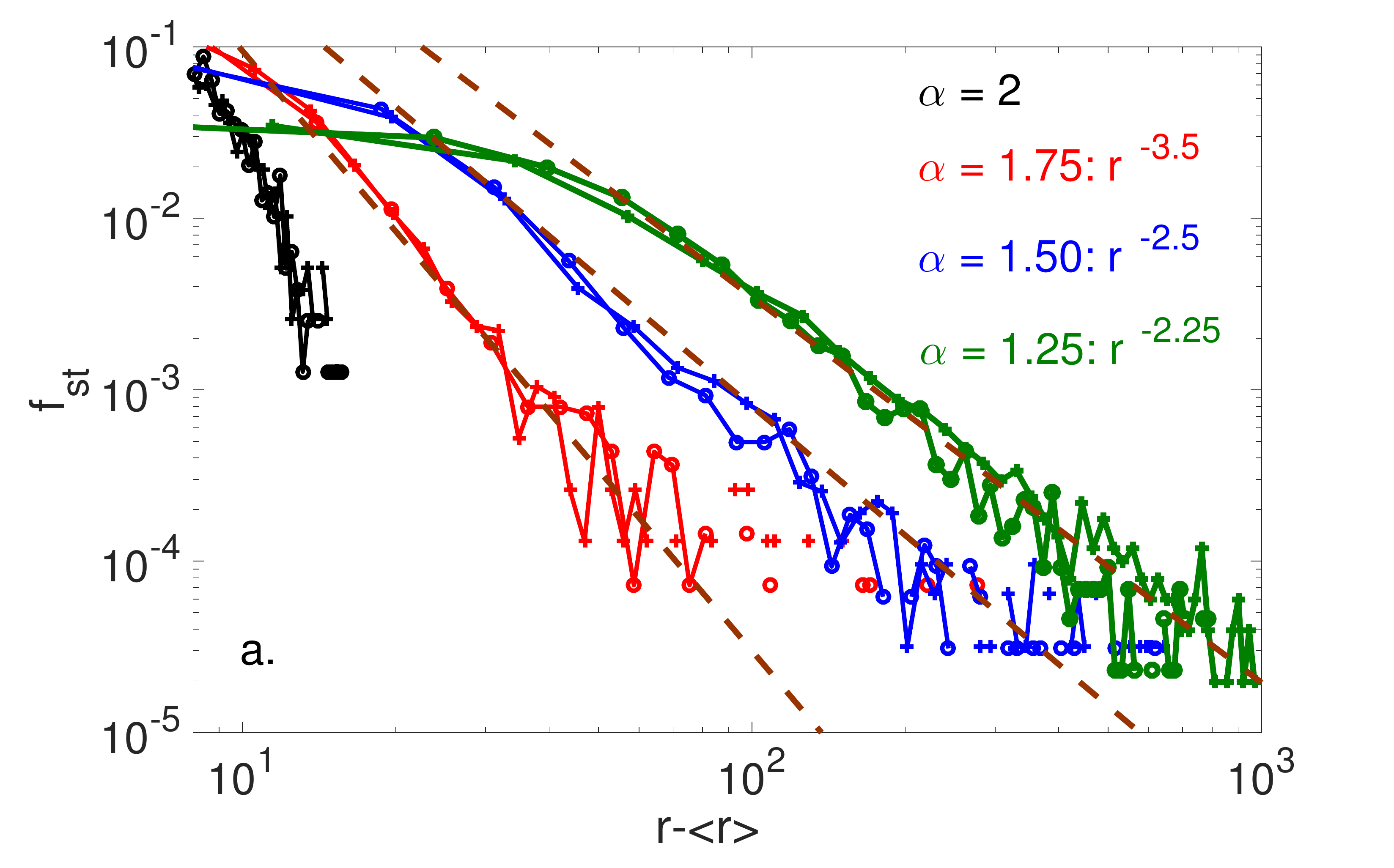} \\
\includegraphics[width=6cm, height=4.cm]{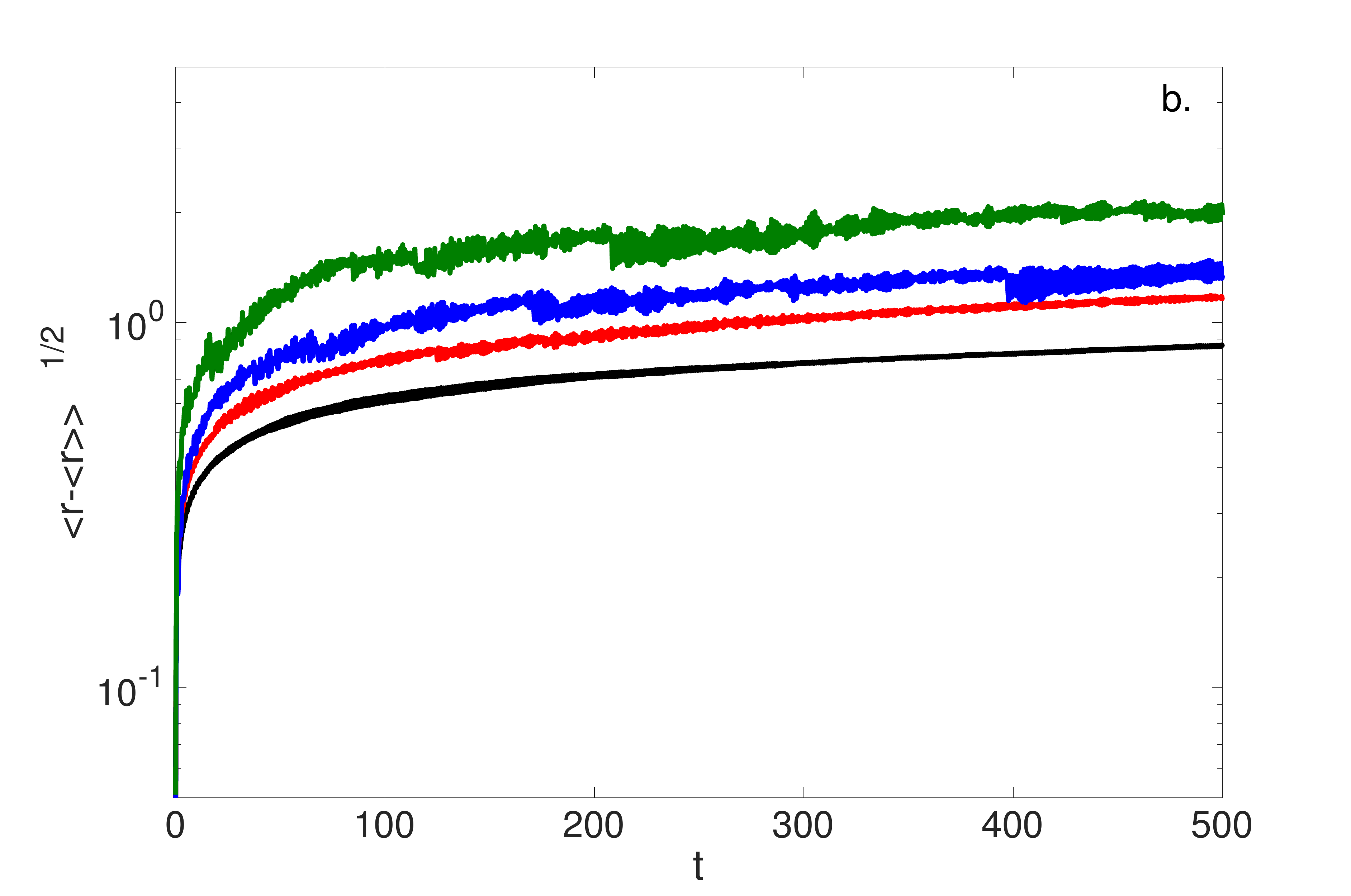} 
\caption{\label{fig6} (a) The steady state PDFs ($f^{st}$) of $r -\langle r\rangle$ where $r=\sqrt{x^2+y^2}/\rho_L(0)$, and (b) the $q$-moment with $q=1/2$ for different values of $\alpha=2$ (black), $1.75$ (red), $1.5$ (blue), and $1.25$ (green). The circle symbols represent the data points on the left of the maximum of $f^{st}$, and cross symbols represent data points on the right. Here, $\epsilon=100$.}
\end{figure}

The PDF of the normalised radial positions, $r -\langle r\rangle$ where $r=\sqrt{x^2+y^2}/\rho_L(0)$ and $\rho_L(0) = v_{\perp}(0)/\Omega_c$ is the particles' Larmor radii at time zero, are shown in Figs. \ref{fig6}(a) and (b) for different values of $\alpha$ in a logarithmic scale. The PDFs of the radial position are influenced by the L\'{e}vy jumps in the velocities with a build up of heavy tails as $\alpha$ is decreased. The $q=1/2$-moments of the particle position show and increase as the stochastic process varies from Gaussian to heavy tailed L\'{e}vy processes. 

\begin{figure}
\includegraphics[width=6cm, height=4.cm]{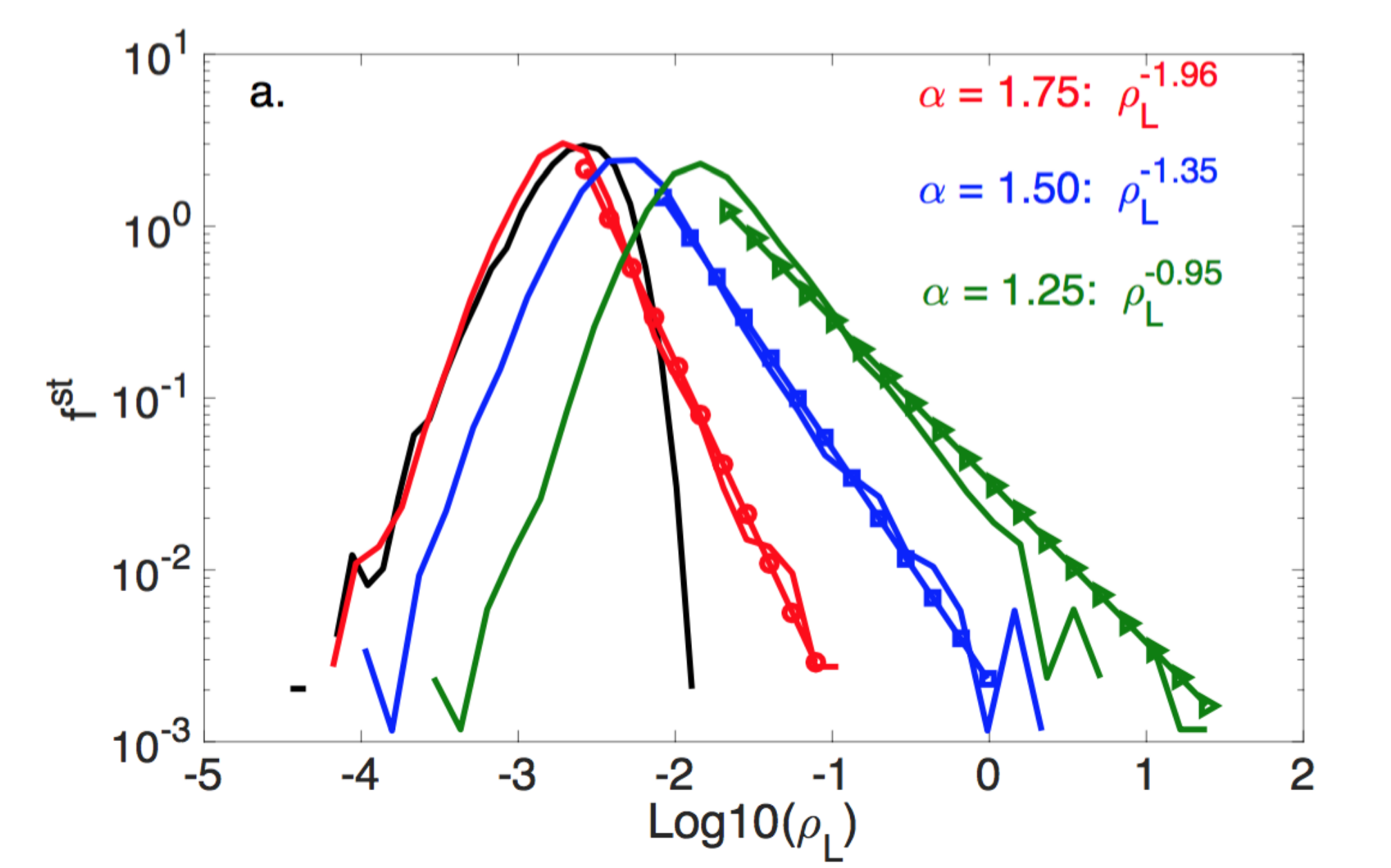}\\
 \includegraphics[width=6cm, height=4.cm]{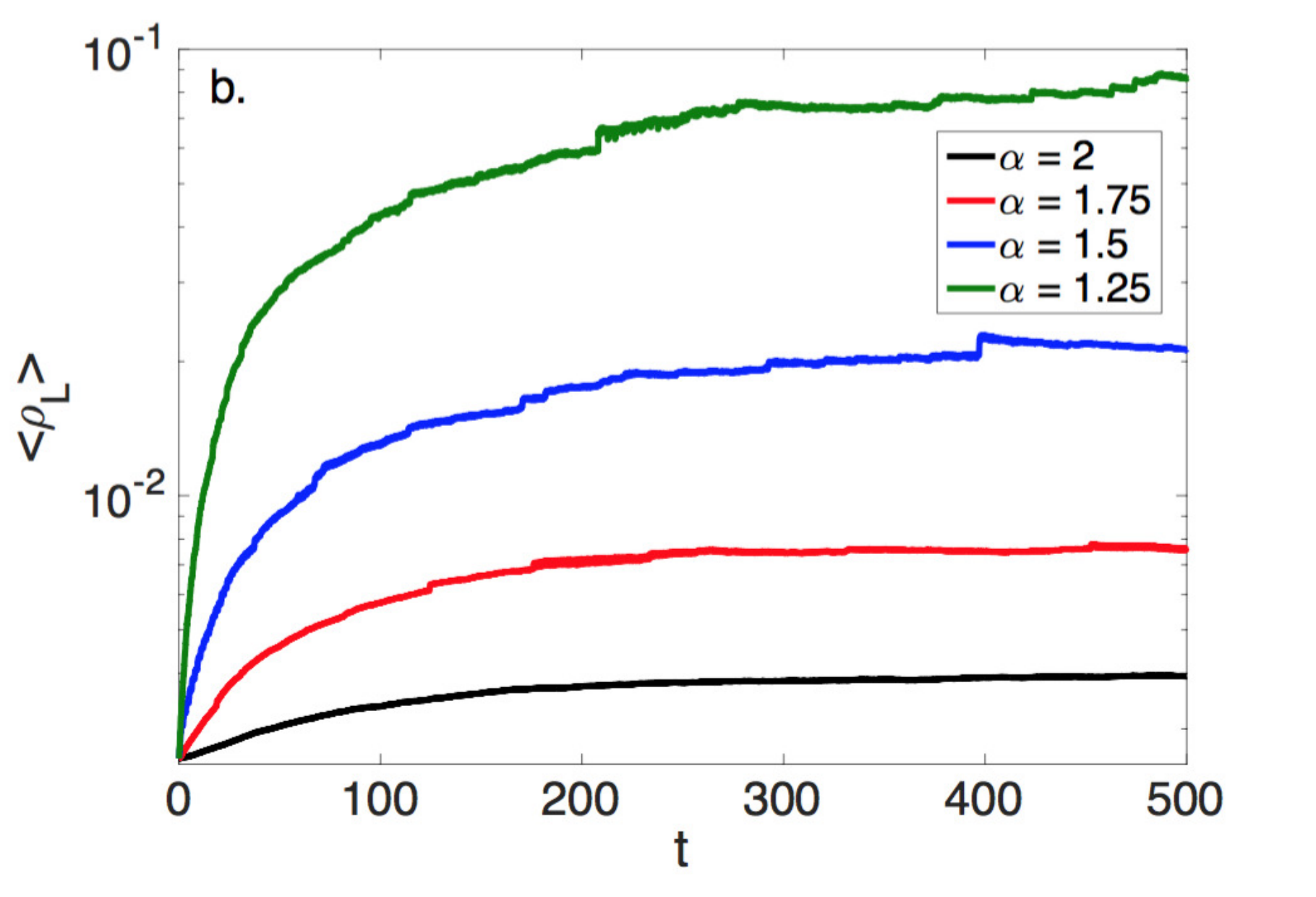}  
\caption{\label{fig7} (a) The steady state PDFs ($f^{st}$) of $Log_{10}(\rho_L)$ of Larmor radii, and linear fits for the decay in the Larmor radii are shown by lines with symbols, and (b) its average as function of time for different values of $\alpha=2$ (black), $1.75$ (red), $1.5$ (blue), and $1.25$ (green). Here, $\epsilon=100$.}
\end{figure}

The PDF of the particles' Larmor radii $(\rho(t) = v_{\perp}(t)/\Omega_c)$, and its average as functions of time, are shown in Figs. \ref{fig7}(a) and (b) for different values of the L\'{e}vy index $\alpha$ in a logarithmic scale. Like in the previous cases, the PDFs change from an exponential decay to a power law decay when the stochastic process is changed from a Gaussian to a L\'{e}vy process. Furthermore, the slope of the power law decay decreases with decrease of $\alpha$ and, as can be seen in the time evolution of the averaged Larmor radius in Fig. \ref{fig6}(b), the converged values are about two orders of magnitude higher in case of the $\alpha=1.25$ as compared to those of the Gaussian with $\alpha=2$. These results suggest that when turbulent electrostatic fluctuation obey non-Gaussian statistics with power law decays, gyro-averaging might be questionable, the guiding centre may not be a valid approximation, and full particle orbits integration should be performed.    

\begin{figure}
\includegraphics[width=6cm, height=4.cm]{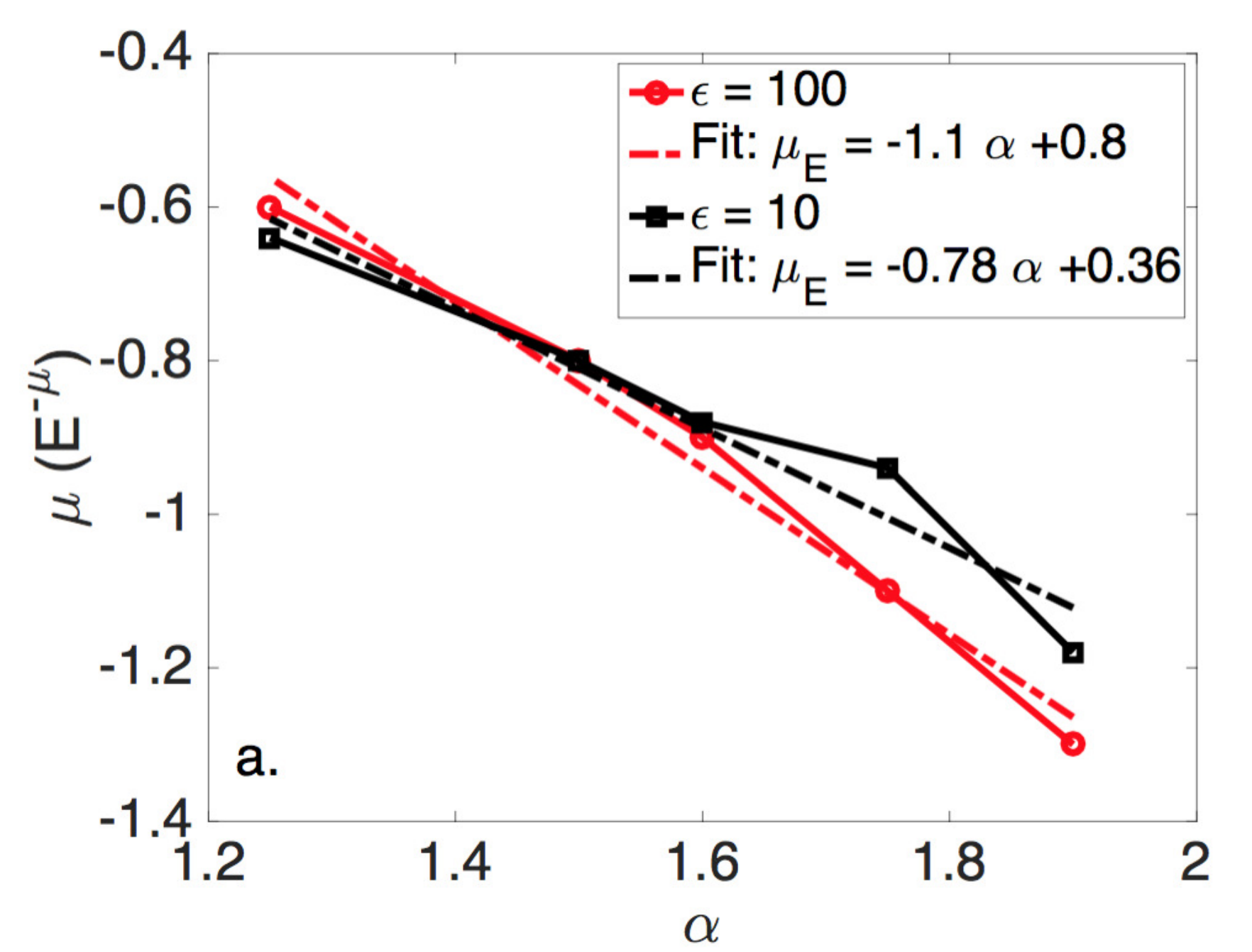}  \\
\includegraphics[width=6cm, height=4.cm]{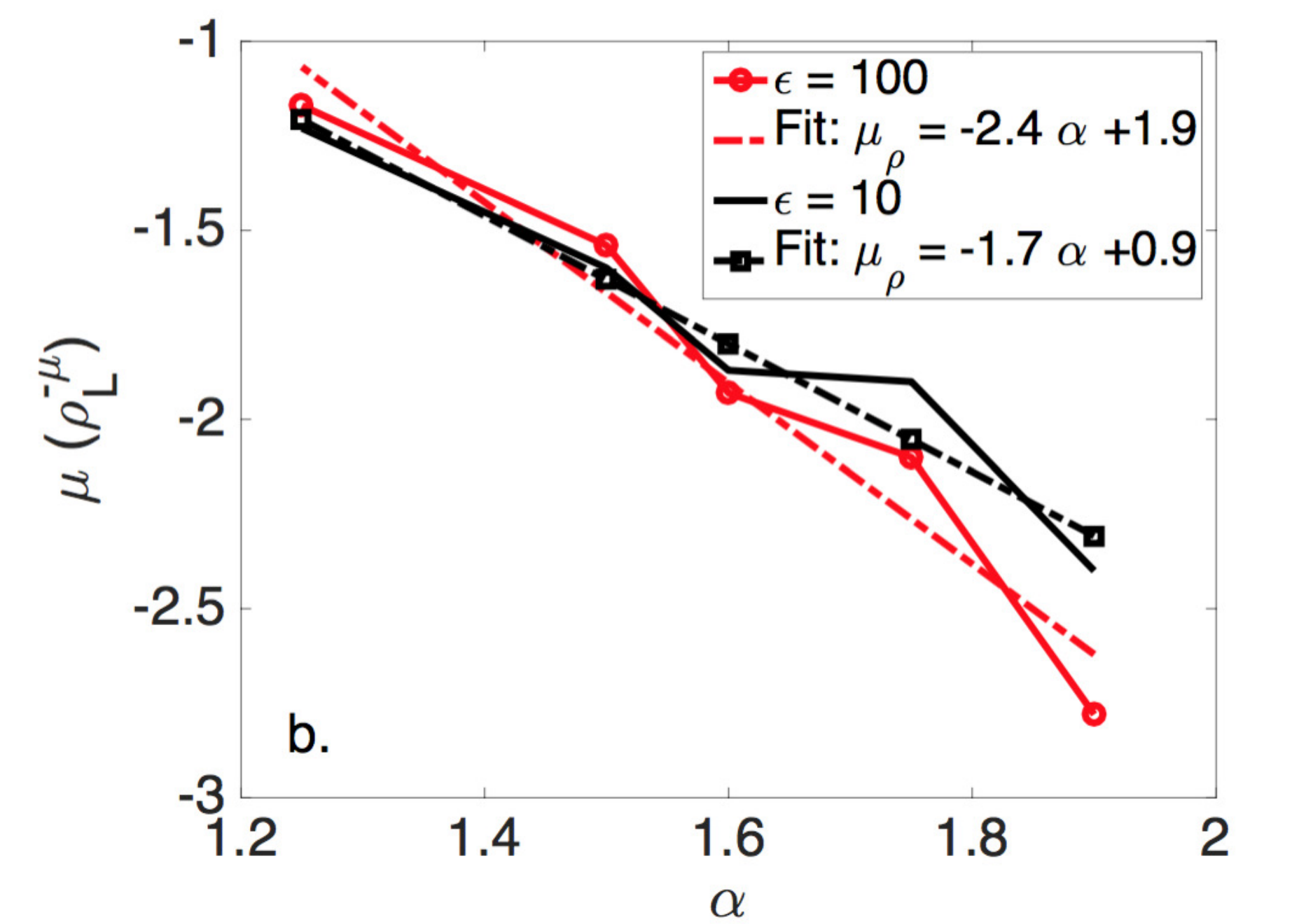}  
\caption{\label{fig8} The linear fits of the power law decay (a) for energy, $\mu_E$, and (b) for the Larmor radius, $\mu_{\rho}$, as functions of the L\'{e}vy index $\alpha$, for different values of $\epsilon=100$ (black line with circle symbols), $\epsilon=10$ (red line with square symbols).}
\end{figure}

The dependence of the power law decay exponents of the energy PDF, $\sim E^{-\mu_E}$, and the  Larmor radius PDF, $\sim \rho_{L}^{-\mu_{\rho}}$, on the L\'{e}vy index $\alpha$ is shown in Fig. \ref{fig8} for $\epsilon=10$ and $100$. The values of the exponents 
$\mu_E$ and $\mu_{\rho}$ where obtained from linear fits of the corresponding PDF in logarithmic scale. 
The results show a close to linear relationship between the exponents $\mu_E$ and $\mu_{\rho}$ and the L\'{e}vy index $\alpha$ with an absolute value of the slope proportional to $\epsilon=\chi/\nu$. 

We have performed Monte Carlo numerical simulations of charged particle motion in the presence of a fluctuating electric field obeying non-Gaussian L\'{e}vy statistics in a constant magnetic field and linear friction modeling the effect of collisional Coulomb drag.  The L\'{e}vy noise was introduced in order to model the effect of non-local transport due to fractional diffusion in velocity space resulting from intermittent electrostatic turbulence. The statistical properties of the velocity moments and energy for various values of the L\'{e}vy index $\alpha$ were investigated, and the role of L\'{e}vy fluctuations on the particles Larmor radii, and the statistical moments of displacements were explored. We observed that as $\alpha$ is decreased, the random walk in energy is strongly influenced by outlier events which result in intermittent behaviour with appearance of L\'{e}vy flights in between periods of small perturbations. The rate and the amplitude of the intermittent jumps in energy increases significantly as $\alpha$ is decreased. The PDFs of the particles' Larmor radii change from an exponential decay to a power law decay when the stochastic electrostatic process is changed from a Gaussian to a L\'{e}vy process. The power law decay decreases with decreasing $\alpha$. This corroborates the findings in Ref. \onlinecite{anderson} that the $q$-moment is an appropriate metric characterizing L\'{e}vy distributed processes. Our findings suggest that when turbulent electrostatic fluctuations  exhibit non-Gaussian L\'{e}vy statistics, gyro-averaging and  guiding centre approximations may not be  fully justified and full particle orbit effects  should be taken into account. The results presented here point out potential limitations of gyro-averaging.  Turbulent plasmas exhibit a very large range of spatio-temporal scales. To overcome the computational challenge that this implies, it is customary to use reduced descriptions based on spatial and/or temporal averaging of degrees of freedom that evolve on small spatial scales and/or fast time scales compared to the macroscopic scales of interest. For example, the extensively used gyro-kinetic models assume $\rho_L / L \ll 1$ where $\rho_L$ is the Larmor radius and $L$ is the tokamak minor radius or a characteristic density gradient scale. However, it is important to keep in mind that in a turbulent plasma the Larmor radius is  a statistical quantity, $\langle \rho_L \rangle$,  (where $\langle \cdot \rangle$ denotes ensemble average) and not an absolute number.  For plasmas in Maxwellian equilibrium this issue might not be critical since the probability density function (PDF) of Larmor radii is sharply peaked around the thermal Larmor radius. However, when the PDF exhibits slowly decaying tails due to a significant number of outliers (i.e., particles with anomalously large Larmor radii) the situation is much less trivial. In particular, in the  case of algebraic decaying PDFs, statistical moments might not exists and as a result in might not be possible to  associate a characteristic scale to the process. The study of scale free stochastic processes has been a topic of significant interest in basic and applied sciences in general and in plasma physics in particular, see for example Refs.~\cite{A,B} and references therein. Our numerical results indicate that when the electrostatic fluctuations follow L\'evy statistics with index $\alpha$, the PDFs of Larmor radii exhibit algebraic decay and this might compromise the meaning of  $\langle \rho_L \rangle$. Formally, if the  PDF of $x \in(0,\infty)$  decays as $ f \sim x^{-\mu}$, then the $n$-th moment, i.e. $\langle x^n \rangle=\int_0^\infty x^n f dx$, will diverge, and thus will not be well-defined, for  $\mu<n+1$.  Based on this, according to Fig.~7, for  $\alpha<1.75$, $\langle \rho_L \rangle$ is  strictly speaking not well-defined. In practice, the mean values might not diverge because as shown in Fig.6-(a) the numerically computed PDFs have a cut-off due to  limited statistical sampling. However, as the case $\alpha=1.25$ in Fig.6-(b) illustrates, the fact that $\mu<2$ implies that the convergence of $\langle \rho_L \rangle$ might be questionable.
Also we would like to note that, in this work as a first step we have limited attention to the study of electrostatic turbulent fluctuations driven by uncorrelated stochastic processes in the absence of memory. However, memory and correlations might play an important role. For example, in Ref.~\cite{del-Castillo-Negrete2005} it was shown that non-Markovian effects are present in  fluid models of plasma turbulent transport and as a consequence, in this case, effective models of particle transport should include both spatial jumps driven by L\'evy processes and memory effects driven by non-Markovian waiting times. On the other hand, the work in Ref.~\cite{Sanchez} showed that correlations play a role on gyro-kinetic turbulent transport in the presence of shear flows and thus, in this case, the proper treatment  requires the use of correlated non-Gaussian random processes. The work presented here could be extended to include memory effects by incorporating non-Markovian statistics in the Monte-Carlo simulation, and also by including correlations using fractional Levy motion models.  Note that doing this would naturally introduce a characteristic time scale into the turbulence fluctuation model, e.g. the correlation time or the memory time scale. A problem of interest would then be to study the dependence of the results on these fluctuation time scales and the gyro-period of the  orbits. These are interesting problems that we plan to address in the future.
\label{discussion}

SM has benefited from a mobility grant funded by the Belgian Federal Science Policy Office and the MSCA of the European Commission (FP7-PEOPLE-COFUND-2008 n$¼$ 246540). DdcN acknowledges support from the Office of Fusion Energy Sciences of the US Department of Energy at Oak Ridge National Laboratory, managed by UT-Battelle, LLC,  for the U.S.Department of Energy under contract DE-AC05-00OR22725.

\end{document}